\documentclass[reprint,superscriptaddress,prr]{revtex4-2}
\usepackage[T1]{fontenc}
\usepackage[latin1]{inputenc}
\usepackage{lmodern}
\expandafter\let\csname equation*\endcsname\relax
\expandafter\let\csname endequation*\endcsname\relax
\usepackage{amsmath}
\usepackage{amssymb,dsfont}
\usepackage{graphicx}
\usepackage{xcolor}
\usepackage{natbib}
\usepackage{bm}
\usepackage{bbold}
\usepackage[mathscr]{euscript}
\usepackage[cal=boondoxo]{mathalfa}
\usepackage{comment}
\renewcommand\footnotemark{}

\usepackage{braket}
\usepackage[normalem]{ulem}
\usepackage{lineno}

\def\ii{{\rm i}}

\def\db{\boldsymbol{\wp}} 
  
\def\rb{{\bf r}}

\def\hge{\hat{\sigma}_{ge}}  
\def\heg{\hat{\sigma}_{eg}}

\def\bra#1{\mathinner{\langle{#1}|}}
\def\ket#1{\mathinner{|{#1}\rangle}}
\def\braket#1{\mathinner{\langle{#1}\rangle}}
\def\jop{\hat{\mathcal{O}}}

\newcommand\varpm{\mathbin{\vcenter{\hbox{%
  \oalign{\hfil$\scriptstyle+$\hfil\cr
          \noalign{\kern-.3ex}
          $\scriptscriptstyle({-})$\cr}%
}}}}
\def\gtau{g^{(2)}(0)}
\def\gthree{g^{(3)}(0)}
\def\dcrit{d_\mathrm{critical}}

\begin{document}

\title{Universality of Dicke superradiance in arrays of quantum emitters}
\author{Stuart J. Masson}
\email{s.j.masson@columbia.edu}
\affiliation{Department of Physics, Columbia University, New York, NY 10027, USA}
\author{Ana Asenjo-Garcia}
\email{ana.asenjo@columbia.edu}
\affiliation{Department of Physics, Columbia University, New York, NY 10027, USA}

\date{\today}

\begin{abstract}

Dicke superradiance is an example of emergence of macroscopic quantum coherence via correlated dissipation. Starting from an initially incoherent state, a collection of excited atoms synchronizes as they decay, generating a macroscopic dipole moment and emitting a short and intense pulse of light. While well understood in cavities, superradiance remains an open problem in extended systems due to the exponential growth of complexity with atom number. Here we show that Dicke superradiance is a universal phenomenon in ordered arrays. We present a theoretical framework -- which circumvents the exponential complexity of the problem -- that allows us to predict the critical distance beyond which Dicke superradiance disappears. This critical distance is highly dependent on the dimensionality and atom number. Our predictions can be tested in state of the art experiments with arrays of neutral atoms, molecules, and solid-state emitters and pave the way towards understanding the role of many-body decay in quantum simulation, metrology, and lasing.

\end{abstract}

\maketitle

\section*{Introduction}

Atoms in close proximity alter each others' radiative environment and collectively interact with light~\cite{Dicke54,Rehler71,Gross82,BenedictBook}.  The ``environment'' for each of the atoms depends on the internal state of all others, which changes in time. For fully-inverted atoms at a single spatial location, this leads to the emission of a short pulse of light that initially rises in intensity, in contrast to the exponential decay of independent atoms. This ``superradiant burst'', or Dicke superradiance, occurs because atoms synchronize as they decay, locking in phase and emitting at an increasing rate. Superradiant bursts have been observed in a variety of dense disordered systems~\cite{Gross82,BenedictBook,Skribanowitz73,Inouye99,Scheibner07,Raino18,Ferioli21PRL}. Dicke superradiance has also been demonstrated in cavities~\cite{Raimond82,Slama07}, where the condition of atoms at a point is emulated by the confinement of the optical field to zero dimensions. In this high-symmetry scenario, atoms are indistinguishable from each other, and can only occupy states that obey a particle-exchange symmetry. This restricts the Hilbert space to permutationally symmetric states, whose number scales linearly with atom number, thus making the dynamical evolution exactly solvable. 

Numerical studies of superradiant emission in extended geometries (of sizes larger than the emission wavelength) have been limited to small numbers of atoms~\cite{Clemens03,Masson20PRL}, small numbers of excitations~\cite{Scully06}, or uniform atomic densities where specific atomic positions are not taken into account~\cite{Friedberg74PRA}. However, recent experimental demonstrations of ordered atomic arrays, via optical tweezers~\cite{Kim16,Endres16,Barredo16,Norcia18,Saskin19,OhlDeMello19} and optical lattices~\cite{Bakr10,Sherson10,Greif16,Kumar18}, open a new world of possibilities, where hundreds of atoms can be placed in almost arbitrary positions. These setups thus demand a new outlook on the problem, which has remained open until now due to the exponential growth of the Hilbert space. In extended systems, particle-exchange symmetry is broken and numerical calculations require a Hilbert space which grows as $2^N$, where $N$ is the atom number.

\begin{figure}[b]
\includegraphics[width=0.45\textwidth]{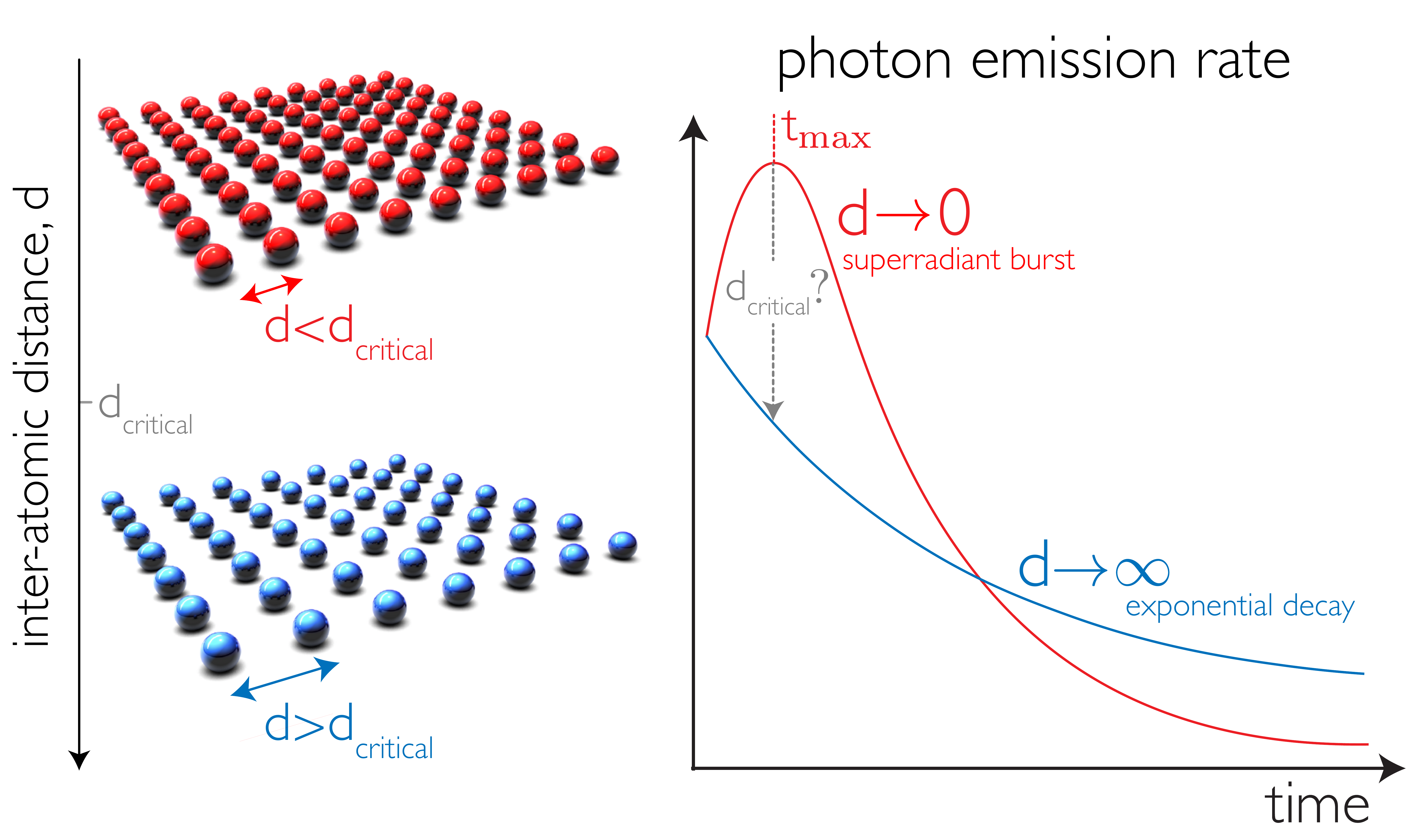}
\caption{Many-body decay is determined by the distance between atoms and the array's dimensionality. Inverted atoms placed at the same location ($d\rightarrow 0$) interact with each other and decay collectively via the emission of a burst of light, with a peak at time $t_\text{max}$. This is the hallmark of Dicke superradiance. In contrast, atoms that are far separated ($d\rightarrow\infty$) emit as single entities, in the form of an exponentially decaying pulse. For extended finite arrays, there is a critical distance at which the crossover between a superradiant burst and monotonically decreasing emission occurs.}\label{figintro}
\end{figure}

Here, we introduce a theoretical framework that scales linearly with atom number and allows us to demonstrate that Dicke superradiant decay generically arises in extended systems, below a critical inter-atomic distance that depends on the dimensionality. We do so by noting that there is no need to compute the full dynamical evolution of the system: the nature of the decay can be deduced from the statistics of the first two emitted photons. We find that as the inter-atomic distance increases, there is a smooth crossover between a superradiant and a monotonically-decreasing emission rate, as shown in Fig.~\ref{figintro}. We obtain an analytical ``minimal condition'' for Dicke superradiance, which is universal and provides a bound on the maximal inter-atomic separation to observe this phenomenon. This enables us to study the role of geometry in the decay of very large atomic arrays, a significant conceptual advance on a decades-old problem.

\section*{Theory}

We first present the theoretical toolbox to describe the dynamics of a collection of atoms interacting via a shared electromagnetic field. We consider $N$ identical two-level atoms of spontaneous emission rate $\Gamma_0$ and transition wavelength $\lambda_0$ placed in free space with arbitrary positions. After tracing out the electromagnetic field using a Born-Markov approximation~\cite{Gruner96,Dung02}, the atomic density matrix $\rho=\ket{\psi}\bra{\psi}$ evolves as
\begin{equation}\label{masterequation}
\dot{\rho} = - \frac{\ii}{\hbar} [\mathcal{H},\rho] + \underbrace{\sum\limits_{\nu=1}^N \frac{\Gamma_\nu}{2} \left( 2\jop_\nu \rho \,\jop_\nu^\dagger - \rho\, \jop_\nu^\dagger\jop_\nu - \jop_\nu^\dagger\jop_\nu \rho \right)}_{\text{dissipative evolution: correlated photon emission}},
\end{equation}
where the Hamiltonian $\mathcal{H}$ describes coherent interactions between atoms and $\{\jop_\nu\}$ are operators that represent how atoms ``jump'' from the excited to the ground state by collectively emitting a photon. Jump operators are found as the eigenstates of the $N\times N$ dissipative interaction matrix $\mathbf{\Gamma}$ with elements $\Gamma^{ij}$, proportional to the  propagator of the electromagnetic field (i.e., the Green's function) between pairs of atoms $i$ and $j$ (see Refs.~\cite{Gruner96,Dung02,Carmichael00,Clemens03} and Supplementary Material). The corresponding eigenvalues provide the jump operator rates $\{\Gamma_\nu\}$, which represent how frequently such a jump occurs. Each of these jump operators imprints a phase in the atoms, and generates a photon with a specific spatial profile in the far field. They thus can be understood as collective ``decay channels'' for the atomic ensemble.

As we demonstrate below, Dicke superradiance is preserved as long as the number of (relevant) decay channels $\nu$ is small. This occurs because dissipative interactions (rather than coherent Hamiltonian dynamics) are responsible for the suppression of superradiance in ordered arrays~\cite{Clemens03,Masson20PRL}. In the paradigmatic example studied by Dicke, where all atoms are exactly at one point, only one of the decay channels is bright (with decay rate $\Gamma_\text{bright}=N\Gamma_0$), while all the others are completely dark (i.e., $\Gamma_\nu=0$). This means that the only possible decay path to the ground state for atoms that are initially excited is through repeated action of the bright operator, which imprints a phase pattern in the atoms that is reinforced in each process of photon emission. Coherence emerges via this dissipative mechanism, which leads to the development of a macroscopic dipole through synchronization and to a rapid release of energy in the form of a superradiant burst. 

In ordered arrays, the number of bright decay channels can be controlled by the inter-atomic distance. In principle, all jump operators are allowed to act. For small lattice constants, their decay rates vary dramatically due to constructive and destructive interference. They can be larger (bright) or smaller (dark) than the single atom decay rate $\Gamma_0$. Extremely dark rates (which are strictly zero in the thermodynamic limit)
emerge for inter-atomic separations below a certain distance that depends on the dimensionality of the array~\cite{Asenjo17PRX}. As the distance grows, the distribution of the decay rates becomes more uniform. This leads to a strong competition between different decay channels, and to decoherence through the randomization of the atomic phases after several emission processes have occurred.

We show here that Dicke superradiance generically occurs in arrays, but only below a critical inter-atomic distance, which can be calculated with a complexity that scales only linearly with system size. For a fixed atom number, the superradiant burst diminishes as the inter-atomic distance increases, eventually being replaced by a monotonically decaying pulse. The crossover between these regimes is marked by an infinitesimally small burst that occurs at $t=0$~\cite{Masson20PRL}. 

\begin{figure}[t!]
    \centering
    \includegraphics[width=0.45\textwidth]{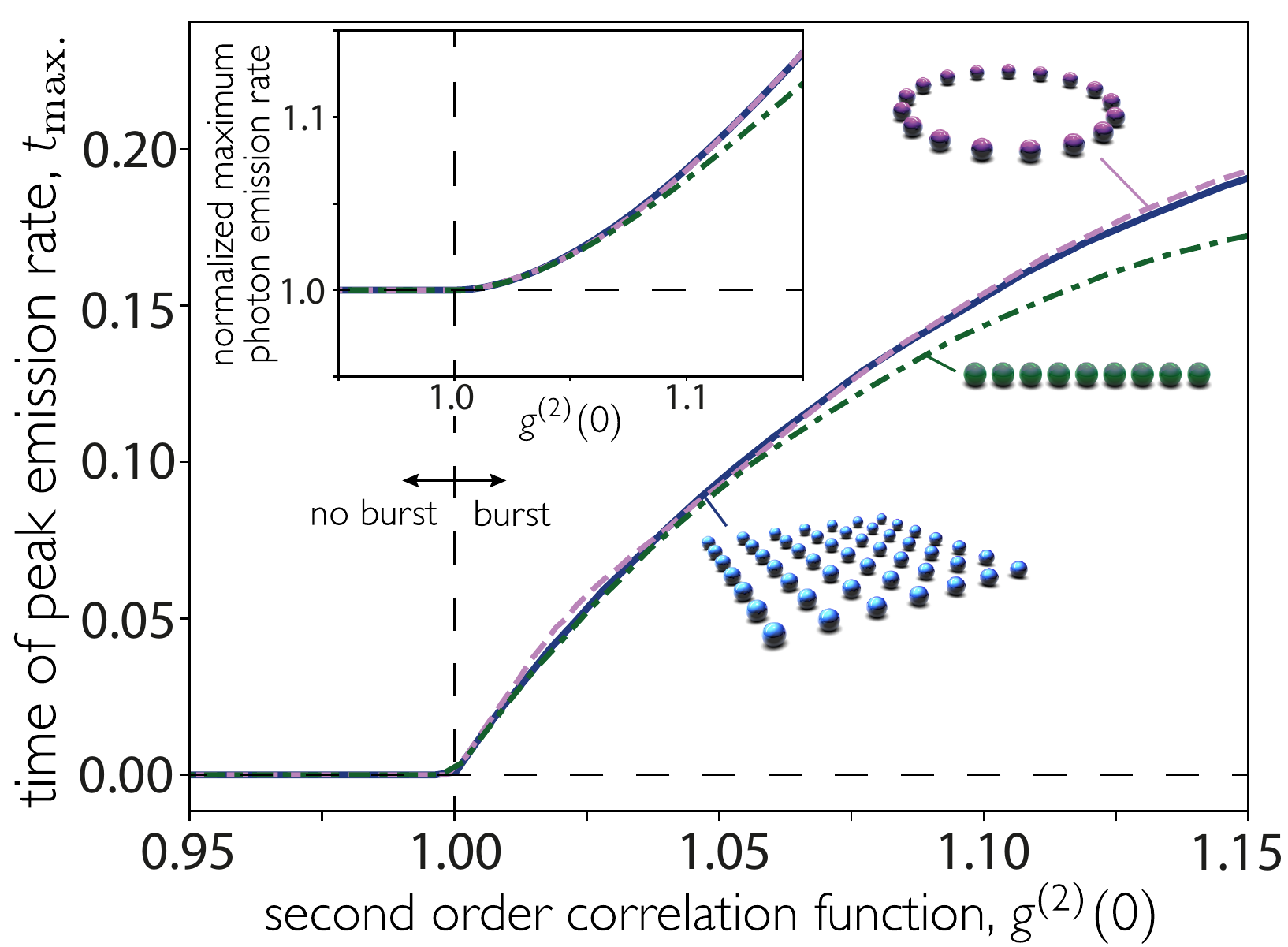}
    \caption{Photon statistics predicts Dicke superradiance. We calculate $g^{(2)}(0)$ (at $t=0$), as an enhanced two-photon emission rate is a pre-requisite for a burst. The time at which the photon rate is maximum ($t_\text{max}$) as a function of the second order correlation function (at $t=0$) shows that $t_\text{max}> 0$ only if $g^{(2)}(0)>1$. Inset: Maximum intensity, normalized by intensity at $t=0$. In both plots, all 9 atoms are initially excited, with polarization perpendicular to the array.
    \label{Fig1}}
\end{figure}

Our key insight is that atomic synchronization occurs immediately or not at all, and thus the nature of the decay can be characterized from early dynamics. In particular, one can predict the presence of a superradiant burst based solely on the statistics of the first two emitted photons. The minimum requirement for a superradiant burst to occur is that the first photon enhances the emission rate of the second. This is captured by the second order correlation function
\begin{equation}
\gtau = \frac{\sum\limits_{\nu,\mu=1}^N \Gamma_\nu\Gamma_\mu \braket{\jop^\dagger_\nu\jop^\dagger_\mu\jop_\mu\jop_\nu}}{\left(\sum\limits_{\nu=1}^N \Gamma_\nu \braket{\jop_\nu^\dagger\jop_\nu}\right)^2},
\end{equation}
where the expectation value is taken at the initial state, i.e., $\ket{\psi(t=0)}=\ket{e}^{\otimes N}$. When this quantity is greater than unity, the decay is characterized as superradiant. Figure~\ref{Fig1} shows the correlation between $\gtau$ and the presence or absence of a burst for small atom numbers, for which we can calculate the full dynamics. As soon as $\gtau>1$, the time of maximum emission deviates from zero (i.e., the burst occurs at a finite time). Moreover, the second order correlation function increases along with the height of the peak of the photon emission rate, and is below unity when the rate is peaked at $t=0$. 

By calculating $\gtau$ analytically (see Supplementary Material), we obtain the minimal condition for Dicke superradiance:
\begin{equation}
\gtau > 1 \;\;\;\; \Leftrightarrow \;\;\;\; \mathrm{Var}\left(\frac{\{\Gamma_\nu\}}{\Gamma_0}\right) > 1,
\end{equation}
where Var is the variance of the decay rates of the jump operators. This expression is exact and universal, and does not involve any assumption about the atomic positions. Small inter-atomic distances maximize the variance of the decay rates, as most jump operators will be dark (with $\Gamma_\nu\simeq 0$) and just a small number of them will be bright (with a large $\Gamma_\nu$).

We note that the complexity of the problem has decreased tremendously: from solving a differential equation in an exponentially-growing Hilbert space, to diagonalizing a matrix whose dimension scales linearly with atom number. This allows one to find the distance at which Dicke superradiance disappears in arbitrary geometries with an extremely large atom number, as all the necessary details are captured in the dissipative interaction matrix $\mathbf{\Gamma}$. Of course, one has to pay a price for this reduction in complexity. As we cannot calculate the full evolution, we can only predict whether a superradiant burst is going to occur or not. Extracting information about the height of the peak or the time at which it will appear requires a different approach~\cite{Robicheaux21PRA_HigherOrderMeanField,Rubies21arxiv}.

\begin{figure}[b!]
    \centering
    \includegraphics[width=0.45\textwidth]{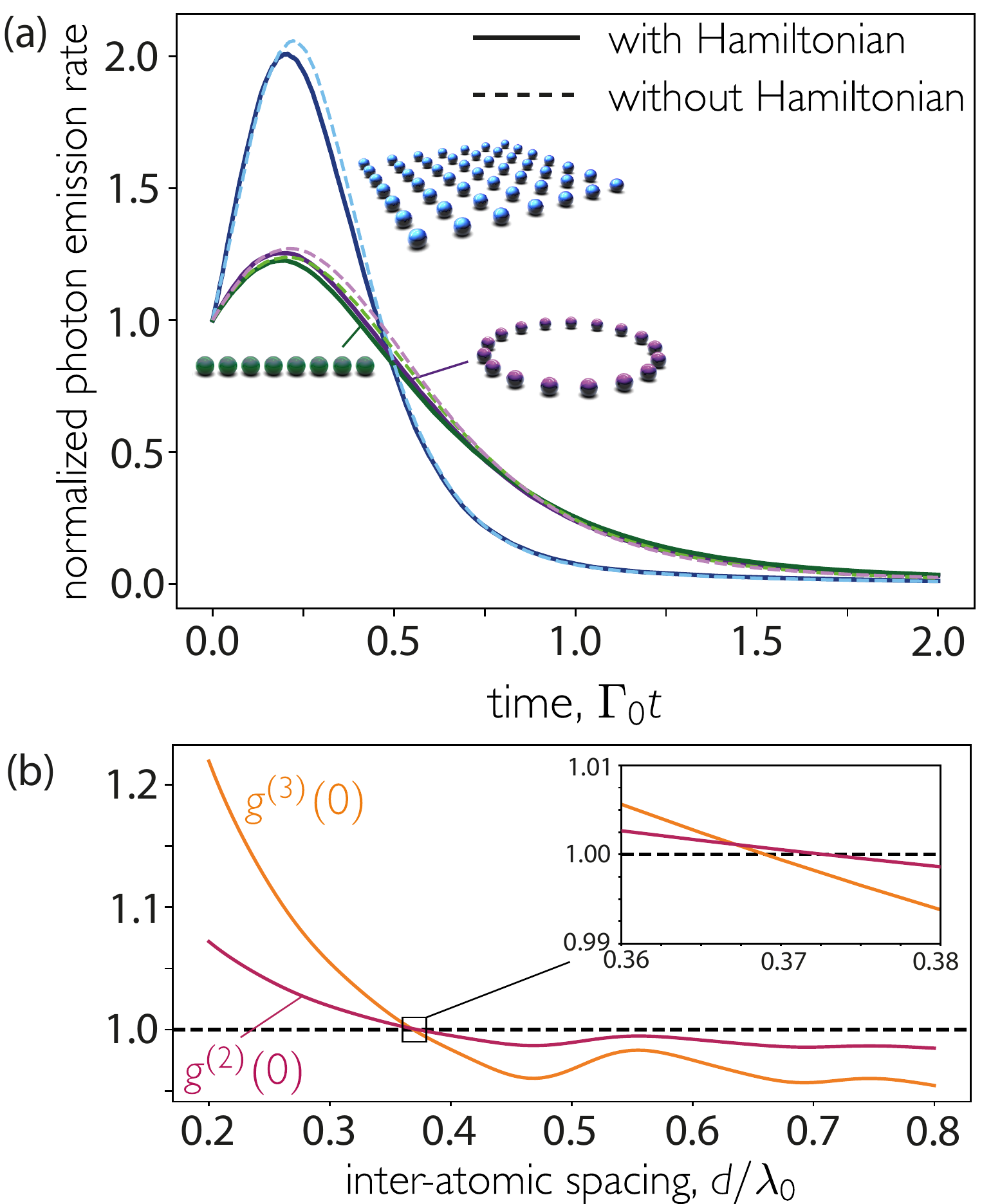}
    \caption{Role of coherent and dissipative evolution in dephasing and suppression of Dicke superradiance. (a) The coherent evolution does not significantly modify the early time dynamics, thus preserving superradiance, as shown by the full evolution of the master equation [i.e., Eq.~(\ref{masterequation})] for 16 initially excited atoms with inter-atomic distance $d=0.1\lambda_0$ arranged in different geometries with and without Hamiltonian interactions. Emission rate is normalized by that at $t=0$. (b) Three-photon decay is never enhanced unless two-photon emission is too, as demonstrated by the second and third order correlation functions, plotted as a function of the inter-atomic separation for a square 2D array of $6\times 6$ atoms. In all cases, atoms are polarized perpendicular to the array.
    \label{Fig2}}
\end{figure}

To prove that the above inequality can be used to characterize Dicke superradiance, we demonstrate that the system does not rephase at later times, either through Hamiltonian action or through further dissipation. First, Hamiltonian dynamics are not significant at early times, as shown in Fig.~\ref{Fig2}(a). Due to the ordered nature of the array, each atom (except those near the boundaries) experiences a similar environment and local dephasing due to Hamiltonian action is thus minimized. To further confirm this point, we consider a time delay between the emission of the first two photons, during which the Hamiltonian acts, and find the Hamiltonian adds a slow dephasing to the atoms but, importantly, does not enhance photon emission (see Supplementary Fig.~\ref{SIFig4}). Second, dissipation into different channels cannot rephase the atoms, as the process is irreversible. With each photon that is emitted, there is one less atom able to emit. To obtain a superradiant burst, the induced atomic correlations that emerge through decay must be large enough to compensate for the loss of emitters, a process that gets harder the more photons have been radiated away. We characterize this through the third order correlation function, which can be analytically calculated (see Supplementary Material). In all geometries considered here, we find that the third photon is never enhanced when the second photon is not. Therefore, further jumps do not rephase the array, as  shown in Fig.~\ref{Fig2}(b), where $g^{(3)}(0)$ drops below unity at a slightly smaller distance than $\gtau$. As anticipated, the second photon is always the last one to lose its stimulated enhancement.

\begin{figure*}[t]
    \centering
    \includegraphics[width=\textwidth]{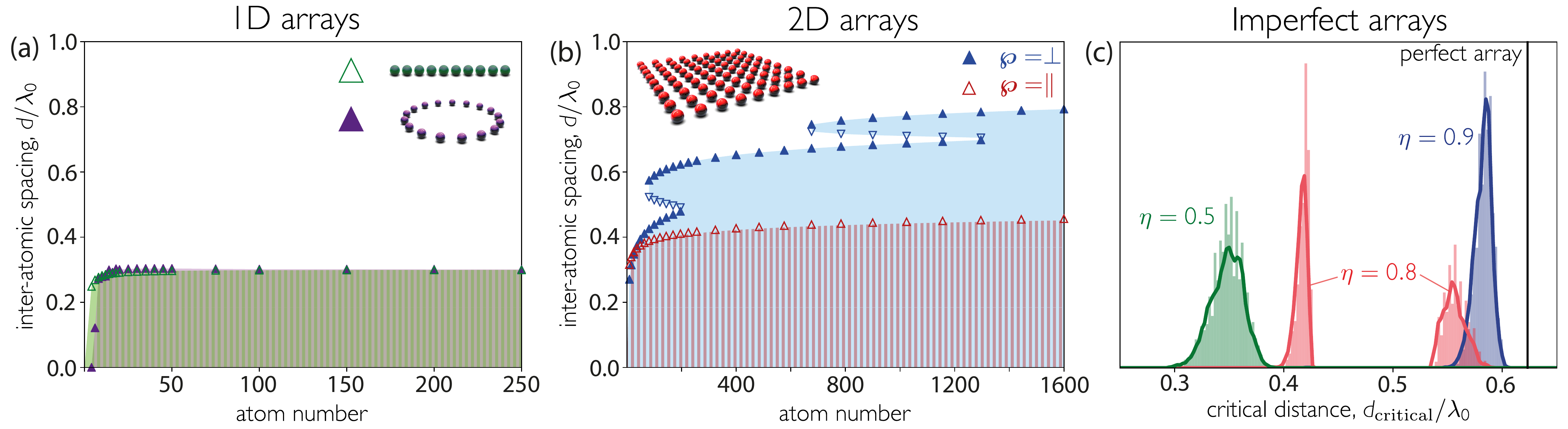}
    \caption{Dicke superradiance is universal and appears (below a critical distance) for arrays of any dimensionality, including imperfectly-filled ones. (a,b)~Boundaries between the burst (colored) and no-burst (white) regions as a function of inter-atomic distance $d$ and atom number for (a)~chains and rings and (b)~square arrays. The crossover occurs where $\gtau=1$. Upward pointing and downward pointing triangles represent points where, with decreasing $d$, $\gtau$ goes above and below unity, respectively.  (c)~Critical distance for different filling fractions $\eta$. The histogram shows 2000 configurations of a $12\times12$ site square array stochastically filled with efficiency $\eta$. Envelopes are calculated as rolling averages. Atoms are polarized (a) parallel to the array (for the ring this implies a spatially-dependent polarization), (b) perpendicular (blue) and parallel (red) to the plane, and (c) perpendicular to the plane.\label{Fig3}}
\end{figure*}

\section*{Results and discussion}

Contrary to the accepted understanding in the literature~\cite{Gross82,BenedictBook}, we find that large chains and rings behave almost identically, as both do not emit a superradiant burst above $\dcrit \approx 0.3\lambda_0$, as shown in Fig.~\ref{Fig3}(a). Despite the ring's particle-exchange symmetry, the difference between the ring and the chain is negligible for large atom number. This is because dephasing is caused by competition between multiple decay channels, which exist regardless of the array topology~\cite{Masson20PRL}. Interactions across the diameter of the ring are very weak, so the exchange symmetry does not matter, as the atoms essentially see the same local environment in both cases. 

Two- and three-dimensional arrays also display Dicke superradiance, at larger inter-atomic separations than those found in chains. Interestingly, the total size of the array is much larger than a wavelength. Figure \ref{Fig3}(b) shows the critical distance for a 2D square array of up to $40\times40$ atoms. In this geometry, the critical distance is not monotonic with the atom number. These sudden variations are due to  ``revivals'' in $\gtau$, which can be seen in Fig.~\ref{Fig2}(b), associated with changes in the distribution of $\set{\Gamma_\nu}$ as the lattice constant hits certain geometric resonances (see Ref.~\cite{Bettles15} and Supplementary Fig.~\ref{SIFig2}). For large arrays (of $N\sim 40\times 40$ atoms), the critical distance is as large as $\dcrit \approx 0.8\lambda_0$ for atoms polarized perpendicular to the array surface, and it seems to continue increasing with atom number, albeit slowly.

Dicke superradiance is due to the dominance of particular decay channels, whose emission is enhanced due to constructive interference. Since the sum of the decay rates is always $N\Gamma_0$ (regardless of the atomic positions), these bright decay channels must be balanced by dark decay channels to maximize the variance. In ordered arrays, the presence of extremely dark channels is explained by energy-momentum mismatch, where some channels correspond to spin waves with wave-vectors outside the light cone~\cite{Zoubi10,Asenjo17PRX}. In 2D arrays, the spin wave with equal phase on all sites, with an in-plane wave-vector $\textbf{k}=0$, emits perpendicular to the array. If the atomic dipole axis points in that direction then emission is forbidden, creating a region of subradiance that persists up to $d < \lambda_0$~\cite{Asenjo17PRX}. Hence, the crossover between superradiant to monotonic decay occurs at much larger distances for atoms with this polarization.  The same phenomenon exists in 3D lattices for any linear polarization axis. Large 2D and 3D lattices both have values of $\dcrit$ well beyond $\lambda_0$~\cite{Robicheaux21PRA_Directional,sierra21arxiv,Rubies21arxiv}. For these higher dimensions, the dominance of certain channels is maintained due to robust constructive inteference between many neighbors, compensated by large numbers of somewhat subradiant, but not perfectly dark, channels.

We demonstrate that Dicke superradiance is robust to imperfections typically found in experiments, such as filling fraction smaller than unity. Figure~\ref{Fig3}(c) shows the bound for stochastically generated $12\times12$ arrays filled with efficiency $\eta$. For $\eta=90\%$, there is a small reduction in the critical distance. However, at $\eta=50\%$, the drop is much larger. This is because the revivals in $\gtau$ are particularly muted by imperfect filling and, at this efficiency, do not breach unity. This phenomenon is also responsible for the splitting of the values of $\dcrit$ at $80\%$ filling efficiency. Superradiance is also robust to position disorder and small imperfections in the initial state (see Supplementary Fig.~3).

Dicke superradiance should thus be observable in experiments with arrays of inter-atomic separation below the critical distance, which are close to being achieved in state-of-the-art setups~\cite{Rui20,Glicenstein20}. It is important to notice that the critical distance does not signal a sharp transition between monotonic decay and superradiance, but instead a smooth crossover. Experimental signatures would be observable well below this bound. Besides atomic tweezer arrays and optical lattices, solid-state emitters hosted in bulk crystals~\cite{Kornher16,Sipahigil16} or in 2D materials~\cite{Palacios17,Proscia18,Li21} are good candidates to observe this physics. These systems can achieve small lattice constants, although they present other difficulties, such as inhomogeneous broadening and non-radiative decay. Nevertheless, Dicke superradiance is robust against these sources of imperfection (see Supplementary Fig.~4).

Superradiance in an extended array is very different from superradiance in a cavity. In the latter, superradiance involves three phenomena simultaneously: a growth in the photon emission rate, a rapid increase of the population of the cavity mode (due to the burst), and an $N^2$-scaling of the radiated intensity peak. These three concepts are not equivalent for extended arrays in free space, and this has experimental consequences. First, in free space, photons are scattered in all directions, and the relevant geometry is not only that of the array, but that of the array together with the detector. In this work, we effectively integrate over all directions, which would correspond to collecting light over a large solid angle. As photon emission after a jump is directional~\cite{Clemens03,Masson20PRL}, the burst is most optimally measured by a detector placed at the location where the far field distribution of the brightest jump operator is maximal. We note that our methods can be extended to  account for ``directional superradiance''. Recent work~\cite{Robicheaux21PRA_Directional} has shown that, unsurprisingly, the critical distance depends on the angular position of the detector. Second, the peak intensity may no longer scale as $N^2$. Finding the exact scaling is numerically challenging as it requires full dynamical evolution, though it should be accessible in experiments. Nevertheless, we speculate that the scaling will depend on the dimensionality and inter-atomic distance, and will be slower than $N^2$ (approaching $N$ for 1D and with a power law whose exponenent increases with dimension).

In conclusion, we have put forward a universal criterion that shines light into the physics of Dicke superradiance in extended systems. We have also demonstrated that Dicke superradiance universally appears in atomic arrays. We have bounded the critical distance that signals the crossover between monotonic decay and a superradiant burst,  which is far larger than previously anticipated (for arrays of dimensionality higher than 1D). This bound is found by diagonalizing a matrix that scales only linearly with atom number. This method bypasses the exponentially growing Hilbert space required for full evolution by simplifying the problem to the statistics of the first two photons, which allows us to predict superradiance for very large arrays. Our approach could potentially be applied to disordered atomic ensembles~\cite{Guerin16,Ferioli21PRX} (where very small inter-atomic distances are achievable, but introduce large Hamiltonian frequency shifts that may need to be accounted for), to other types of Markovian electromagnetic reservoirs, such as nanophotonic structures~\cite{Goban15,Solano17} (by simply changing the Green's function~\cite{Chang13}), and to emitters with more complex internal or hyperfine structure~\cite{Hebenstreit17,Asenjo19,PineiroOrioli22}. 

The understanding of many-body decay provided by our work is critical for developing robust and scalable quantum applications, ranging from quantum computing and simulation to metrology and lasing. In particular, our work is relevant for Rydberg atom quantum simulators~\cite{Labuhn16,Bernien17,Kim18}, where Rydberg states may decay via long-wavelength transitions. These decay paths may be superradiantly enhanced at short distances~\cite{Wang07,Goldschmidt16}. Atomic arrays are also used in state-of-the-art atomic clocks and other precision measurement experiments~\cite{Bothwell19,Norcia19}. As such systems shrink, it is crucial to understand the impact of collective dissipation. Finally, controlling the light emitted by an atomic array enables its use as an optical source. We have demonstrated that geometry can be used to alter the collective optical properties of the array and shape the temporal profile and statistics of the emitted light. This presents the opportunity to use atomic arrays to produce directional single photons~\cite{Holzinger21}, correlated photons~\cite{Masson20PRL}, or superradiant lasers~\cite{Bohnet12}. Alternatively, measurement of the emitted light provides a window into the complex evolution of the atomic system; and directional detection may enable heralded production of many-body entangled dark states.

\textbf{Acknowledgments --} We are grateful to L. A. Orozco, I. Ferrier-Barbut, A. Browaeys, D. E. Chang, M. Lipson, and E. Sierra for discussions. Research was supported by Programmable Quantum Materials, an Energy Frontier Research Center funded by the U.S. Department of Energy (DOE), Office of Science, Basic Energy Sciences (BES).
We acknowledge computing resources from Columbia University's Shared Research Computing Facility project, which is supported by NIH Research Facility Improvement Grant 1G20RR030893-01, and associated funds from the New York State Empire State Development, Division of Science Technology and Innovation (NYSTAR) Contract C090171, both awarded April 15, 2010.

\textbf{Author contributions --} The numerical analysis was carried out by S. J. M. S. J. M. and A. A.-G. contributed to the development of theoretical ideas and tools, and to the writing of the manuscript. 

\textbf{Competing interests --} The authors declare no competing interests.

\textbf{Data availability --} All data in this manuscript is available upon reasonable request.

\textbf{Code availability --} All code used in this manuscript is available upon reasonable request.

\newpage

\onecolumngrid

\setcounter{figure}{0}  
\renewcommand{\figurename}{Supplementary figure}

\section*{Methods and Supplementary Material}

\subsection*{Atom-atom interactions}

We consider $N$ two-level atoms of resonance frequency $\omega_0$ and spontaneous emission rate $\Gamma_0$ in free space at positions $\set{\rb_i}$. After tracing out the electromagnetic field using a Born-Markov approximation~\cite{Gruner96,Dung02}, the atomic density matrix $\rho$ evolves as
\begin{equation}
\dot{\rho} = - \frac{\ii}{\hbar} [\mathcal{H},\rho] + \sum\limits_{i,j=1}^N \frac{\Gamma_{ij}}{2} \left( 2\hge^i\rho\heg^j  - \rho\, \heg^j\hge^i - \heg^j\hge^i \rho \right),
\end{equation}
where $\hat{\sigma}_{ge}^i = \ket{g_i} \bra{e_i}$ is the atomic coherence operator, $\ket{e_i}$ and $\ket{g_i}$ are the excited and ground states of the $i$th atom, and the Hamiltonian reads
\begin{equation}\label{hamiltonian}
\mathcal{H} = \hbar\sum_{i=1}^N\omega_0\hat{\sigma}_{ee}^i + \hbar\sum_{i,j=1}^N J^{ij}\hat{\sigma}_{eg}^i\hat{\sigma}_{ge}^j.
\end{equation}
The coherent and dissipative interaction rates between atoms $i$ and $j$ are given by~\cite{Carmichael00,Clemens03}
\begin{equation}\label{shiftrate}
J^{ij} - \ii\frac{\Gamma^{ij}}{2} =-\frac{\mu_0\omega_0^2}{\hbar}\,\db^*\cdot \mathbf{G}_0(\rb_i,\rb_j,\omega_0)\cdot\db,
\end{equation}
where $\db$ is the dipole matrix element of the atomic transition and $\mathbf{G}_0(\rb_i,\rb_j,\omega_0)$ is the propagator of the electromagnetic field between atomic positions $\rb_i$ and $\rb_j$~\cite{Gruner96,Dung02}
\begin{align}
    \mathbf{G}_0(\rb_{ij},\omega_0)&=\frac{e^{\ii k_0r_{ij}}}{4\pi k_0^2r_{ij}^3}\left[(k_0^2r_{ij}^2+\ii k_0r_{ij}-1)\mathds{1}  +(-k_0^2r_{ij}^2-3\ii k_0r_{ij}+3)\frac{\rb_{ij}\otimes\rb_{ij}}{r_{ij}^2}\right]
\end{align}
where $\rb_{ij}=\rb_i-\rb_j$ and $r_{ij} = |\rb_{ij}|$. The dissipative interactions can be recast in terms of jump operators, $\set{\jop_\nu}$, found as the $N$ eigenvectors of the matrix $\mathbf{\Gamma}$ with elements $\Gamma_{ij}$. The decay rates, $\set{\Gamma_\nu}$, are found as the corresponding eigenvalues. The atomic master equation thus reads
\begin{equation}
\dot{\rho} = - \frac{\ii}{\hbar} [\mathcal{H},\rho] + \sum\limits_{\nu=1}^N \frac{\Gamma_\nu}{2} \left( 2\jop_\nu \rho \,\jop_\nu^\dagger - \rho\, \jop_\nu^\dagger\jop_\nu - \jop_\nu^\dagger\jop_\nu \rho \right).
\end{equation}
The jump operators are generically a superposition of lowering operators and can be expanded as
\begin{equation}
\jop_\nu = \sum\limits_{i=1}^N \alpha_{\nu,i} \hge^i,\;\;\;\;\mathrm{where}\;\;\;\;\sum\limits_{i=1}^N \alpha_{\nu,i}^*\alpha_{\mu,i} = \delta_{\nu\mu}\;\;\;\;\mathrm{and}\;\;\;\; \sum\limits_{\nu=1}^N \Gamma_\nu |\alpha_{\nu,i}|^2 = \Gamma_0.\label{opform}
\end{equation}
In the above expression, $\delta_{\mu\nu}$ is the Kronecker delta function and $\alpha_{\nu,i}$ is the spatial profile of the $\nu-$jump operator. The total photon emission rate is calculated as
\begin{equation}
R = \sum\limits_{\nu=1}^N \Gamma_\nu \braket{\jop^\dagger_\nu\jop_\nu}.
\end{equation}

\subsection*{Derivation of the second order correlation function $\gtau$}

The second order correlation function is calculated as
\begin{equation}
\gtau = \frac{\sum\limits_{\nu,\mu=1}^N \Gamma_\nu\Gamma_\mu \braket{\jop_\nu^\dagger \jop_\mu^\dagger \jop_\mu \jop_\nu}}{\left(\sum\limits_{\nu=1}^N \Gamma_\nu \braket{\jop_\nu^\dagger \jop_\nu}\right)^2}, \label{g2tau0}
\end{equation}
where the expectation value is taken on the fully excited state $|e\rangle^{\otimes N}$, which is the initial state of the system. Substituting in the form of the operators, as shown in Eq.~\eqref{opform}, one finds
\begin{equation}
\gtau = \frac{\sum\limits_{\nu,\mu=1}^N \Gamma_\nu\Gamma_\mu \sum\limits_{i,j,l,m=1}^N\alpha_{\nu,i}^* \alpha_{\mu,j}^*\alpha_{\mu,l}\alpha_{\nu,m} \braket{\heg^i\heg^j\hge^l\hge^m}}{\left(\sum\limits_{\nu=1}^N \Gamma_\nu \sum\limits_{i,j=1}^N \alpha_{\nu,i}^*\alpha_{\nu,j} \braket{\heg^i\hge^j}\right)^2}. \label{g2tauop}
\end{equation}
On the fully excited state, these expectation values are evaluated as
\begin{align}
\braket{\heg^i\hge^j} = \delta_{ij},\;\;\;\;\;\;\;\;\braket{\heg^i\heg^j\hge^l\hge^m} = \left( \delta_{im}\delta_{jl} + \delta_{il}\delta_{jm}\right)\left(1 - \delta_{ij}\right).
\end{align}
Therefore,
\begin{align}
\gtau &= \frac{\sum\limits_{\nu,\mu=1}^N \Gamma_\nu\Gamma_\mu \left(\sum\limits_{i,j=1}^N |\alpha_{\nu,i}|^2 |\alpha_{\mu,j}|^2 + \sum\limits_{i,j=1}^N\alpha_{\nu,i}^* \alpha_{\mu,j}^*\alpha_{\mu,i}\alpha_{\nu,j} - 2\sum\limits_{i=1}^N|\alpha_{\nu,i}|^2|\alpha_{\mu,i}|^2\right)}{\left(\sum\limits_{\nu=1}^N \Gamma_\nu \sum\limits_{i=1}^N |\alpha_{\nu,i}|^2\right)^2} \notag\\
&= \frac{\sum\limits_{\nu,\mu=1}^N \Gamma_\nu\Gamma_\mu \left[\left(\sum\limits_{i=1}^N |\alpha_{\nu,i}|^2\right)\left( \sum\limits_{j=1}^N|\alpha_{\mu,j}|^2 \right) + \left(\sum\limits_{i=1}^N\alpha_{\nu,i}^* \alpha_{\mu,i}\right)\left(\sum\limits_{j=1}^N\alpha_{\mu,j}^*\alpha_{\nu,j}\right) - 2\sum\limits_{i=1}^N|\alpha_{\nu,i}|^2|\alpha_{\mu,i}|^2\right]}{N^2\Gamma_0^2}\notag\\
&= \frac{\sum\limits_{\nu,\mu=1}^N \Gamma_\nu\Gamma_\mu \left[1 + \delta_{\nu\mu} - \sum\limits_{i=1}^N2|\alpha_{\nu,i}|^2|\alpha_{\mu,i}|^2\right]}{N^2\Gamma_0^2} = \frac{N^2\Gamma_0^2 + \sum\limits_{\nu=1}^N \Gamma_\nu^2 - 2\sum\limits_{i=1}^N\left(\sum\limits_{\nu=1}^N \Gamma_\nu|\alpha_{\nu,i}|^2\right)\left(\sum\limits_{\mu=1}^N \Gamma_\mu|\alpha_{\mu,i}|^2\right)}{N^2\Gamma_0^2} \notag\\
&= 1 + \sum\limits_{\nu=1}^N \left( \frac{\Gamma_\nu}{N\Gamma_0} \right)^2 - \frac{2}{N}\equiv 1 + \frac{1}{N}\left[\mathrm{Var}\left(\frac{\left\{\Gamma_\nu\right\} }{ \Gamma_0}\right) - 1\right].
\end{align}

\subsection*{Derivation of the third order correlation function $\gthree$}

The third order correlation function is calculated as
\begin{align}
\gthree &= \frac{\sum\limits_{\nu,\mu,\chi=1}^N \Gamma_\nu\Gamma_\mu\Gamma_\chi \braket{\jop_\nu^\dagger \jop_\mu^\dagger \jop_\chi^\dagger \jop_\chi \jop_\mu \jop_\nu}}{\left(\sum\limits_{\nu=1}^N \Gamma_\nu \braket{\jop_\nu^\dagger \jop_\nu}\right)^3}. \notag\\
&= \frac{\sum\limits_{\nu,\mu,\chi=1}^N \Gamma_\nu\Gamma_\mu\Gamma_\chi \sum\limits_{i,j,l,m,n,p=1}^N \alpha^*_{\nu,i}\alpha^*_{\mu,j}\alpha^*_{\chi,l}\alpha_{\chi,m}\alpha_{\mu,n}\alpha_{\nu,p}\braket{\heg^i\heg^j\heg^l\hge^m\hge^n\hge^p}}{\left(\sum\limits_{\nu=1}^N \Gamma_\nu \sum\limits_{i,j=1}^N \alpha_{\nu,i}^*\alpha_{\nu,j} \braket{\heg^i\hge^j}\right)^3}\label{gthree}
\end{align}
For the fully-excited state, the denominator is $\left(N\Gamma_0\right)^3$. For the numerator, the expectation value is
\begin{align}
\braket{\heg^i\heg^j\heg^l\hge^m\hge^n\hge^p} &= \left[ \delta_{ip}\left(\delta_{jn}\delta_{lm} + \delta_{jm}\delta_{ln}\right) + \delta_{in}\left(\delta_{jp}\delta_{lm} + \delta_{jm}\delta_{lp}\right) + \delta_{im}\left(\delta_{jp}\delta_{ln} + \delta_{jn}\delta_{lp} \right) \right] \notag
\\&\;\;\;\;\;\;\;\;\;\;\;\;\times \left(1 - \delta_{ij} - \delta_{il} - \delta_{jl} + 2\delta_{ij}\delta_{il}\right).
\end{align}
Using the same relations as above, we calculate the value of $\gthree$ as
\begin{align}
\gthree &= \frac{1}{N^3\Gamma_0^3} \sum\limits_{\nu=1}^N\sum\limits_{\mu=1}^N\sum\limits_{\chi=1}^N \Gamma_\nu \Gamma_\mu \Gamma_\chi \left( 1 + 2\delta_{\nu\mu\chi} + \delta_{\nu\mu}+ \delta_{\nu\chi}+ \delta_{\mu\chi} + 12\sum\limits_{i=1}^N |\alpha_{\nu,i}|^2|\alpha_{\mu,i}|^2|\alpha_{\chi,i}|^2 \color{white}\right)\color{black}\notag\\
&- 2\sum\limits_{i=1}^N |\alpha_{\nu,i}|^2 |\alpha_{\chi,i}|^2  - 2\sum\limits_{i=1}^N  |\alpha_{\nu,i}|^2|\alpha_{\mu,i}|^2  - 2\sum\limits_{i=1}^N |\alpha_{\mu,i}|^2|\alpha_{\chi,i}|^2 \notag\\
&\color{white}\left(\color{black} - 4\delta_{\nu\chi}\sum\limits_{i=1}^N |\alpha_{\nu,i}|^2|\alpha_{\mu,i}|^2 - 4\delta_{\nu\mu} \sum\limits_{i=1}^N  |\alpha_{\nu,i}|^2|\alpha_{\chi,i}|^2- 4\delta_{\mu\chi} \sum\limits_{i=1}^N  |\alpha_{\nu,i}|^2 |\alpha_{\mu,i}|^2 \right) \notag \\
&= \frac{1}{N^3\Gamma_0^3} \left( N^3\Gamma_0^3 + 2\sum\limits_{\nu=1}^N \Gamma_\nu^3 + 3N\Gamma_0\sum\limits_{\nu=1}^N  \Gamma_\nu^2 + 12N\Gamma_0^3 - 6N^2\Gamma_0^3  - 12\Gamma_0\sum\limits_{\nu=1}^N  \Gamma_\nu^2 \right) \notag \\
&= 1 + 2\sum\limits_{\nu=1}^N\left(\frac{ \Gamma_\nu}{N\Gamma_0}\right)^3 + \left(3 - \frac{12}{N}\right)\sum\limits_{\nu=1}^N\left(\frac{  \Gamma_\nu}{N\Gamma_0}\right)^2 + \frac{12}{N^2} - \frac{6}{N}.
\end{align}

\section*{Decay rates as a function of distance in ordered arrays}

Generally, the variance of the eigenvalues $\{\Gamma_{\nu}\}$ increases with decreasing inter-atomic distances. However, this is not always strictly true. At some specific distances, there are geometric resonances that cause the decay rates to experience sudden changes~\cite{Bettles15,Bettles16PRL,Kramer16,Javanainen19}, which leads to an increase in the variance, as shown in Supplementary Figure~\ref{SIFig2}. These resonances are associated with far-field contributions to the interaction, and occur because certain decay channels become significantly brighter due to constructive interference. In 1D, the first revival occurs at $d=\lambda_0/2$. Extremely subradiant states do not exist for this distance, and thus this revival is not enough to enhance two-photon emission and superradiance. In 2D, for atoms polarized perpendicular to the surface, revivals occur at $d=\lambda_0/2$ and $d=\lambda_0/\sqrt{2}$. For these distances in 2D there are subradiant states. The revivals are strong enough to cause superradiance, leading to the non monotonic behavior of the critical distance with atom number observed in Fig.~4(b) in the main text. For atoms with polarization in the plane, far-field emission in the plane is forbidden in the direction that coincides with that of the polarization, greatly quenching the revivals.

\begin{figure}[h!]
\begin{center}
\includegraphics[width=0.7\textwidth]{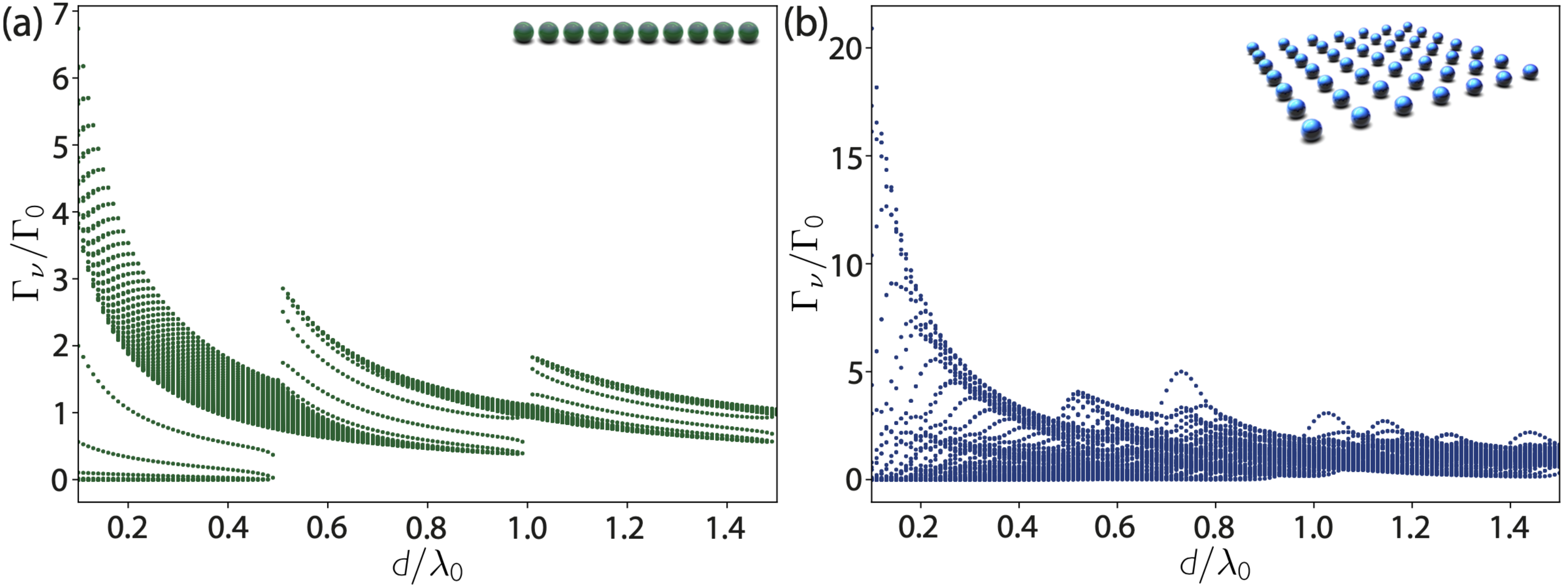}
\end{center}
\caption{Operator decay rates for 100 atoms arranged in a (a) chain and (b) $10\times10$ square array. Atoms are polarized (a) parallel to the array and (b) perpendicular to the array.}\label{SIFig2}
\end{figure}

\section*{Role of Hamiltonian interactions in dephasing}

\begin{figure}[h!]
\includegraphics[width=\textwidth]{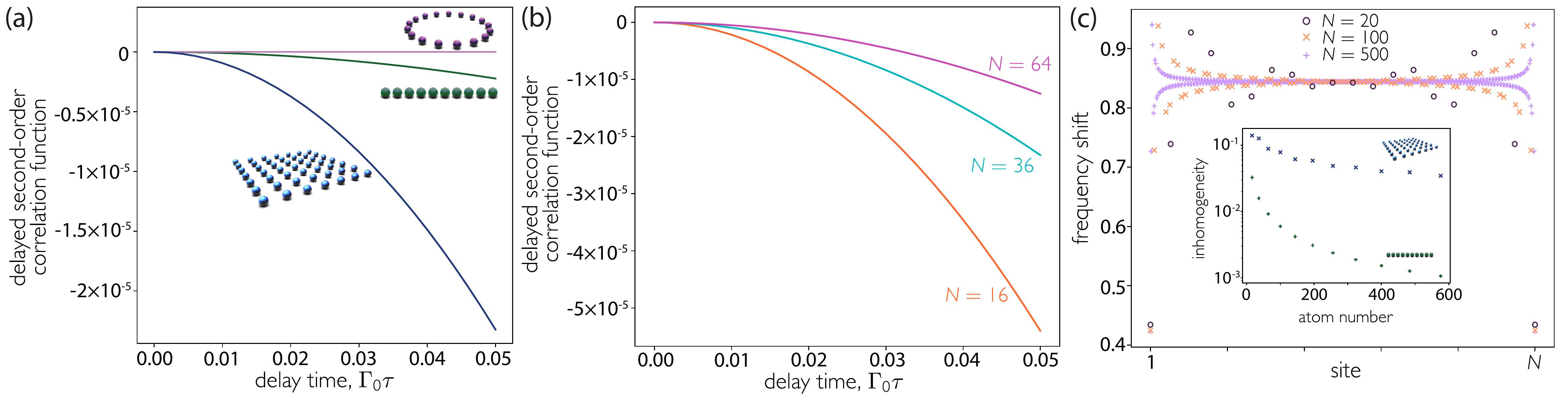}
\caption{Impact of Hamiltonian interactions on the second order correlation function. (a,b) $g^{(2)}(\tau)/g^{(2)}(0) - 1$ is plotted as a function of delay time, showing that the Hamiltonian strictly causes dephasing, or, in the case of the ring, does not have any impact. In (a), calculations are made for different shaped arrays of 36 atoms. In (b), calculations are made for a square array of different number of atoms. In all cases, calculations are made at critical distance and the polarization axis is perpendicular to the array. (c) Frequency shifts for each atom in a linear chain. Inset shows the scaling with atom number of the variance of the frequency shifts normalized by the mean. In both plots, $d/\lambda_0=0.25$.}\label{SIFig4}
\end{figure}

We consider the role of the Hamiltonian by considering a delay time between the first two photons and comparing to the case without a delay. We calculate
\begin{equation}
\frac{g^{(2)}(\tau)}{g^{(2)}(0)} = \frac{\sum\limits_{\nu,\mu=1}^N \Gamma_\nu\Gamma_\mu \braket{\jop_\nu^\dagger \mathrm{e}^{i\mathcal{H}\tau} \jop_\mu^\dagger \jop_\mu \mathrm{e}^{-i\mathcal{H}\tau} \jop_\nu}}{\sum\limits_{\nu,\mu=1}^N \Gamma_\nu\Gamma_\mu \braket{\jop_\nu^\dagger \jop_\mu^\dagger \jop_\mu \jop_\nu}}
\end{equation}
on the fully excited state. This is shown in Supp. Figure~\ref{SIFig4}(a) at the critical distance for different arrays. We note that the Hamiltonian causes very slow dephasing in the case of a linear or square array, and has no impact on the ring array. Calculations show that mixing due to non-measurement introduces an additional (but smaller) dephasing.

The dephasing is reduced with $N$, as shown in Supplementary Figure~\ref{SIFig4}(b) at the critical distance. Hamiltonian dephasing is primarily due to inhomogeneous (i.e., local) frequency shifts caused by interactions~\cite{Friedberg72}. With increasing $N$, $\dcrit$ increases, such that interactions are reduced at the critical distance and dephasing is reduced. Furthermore, atoms that see the similar local environment have similar shifts. This means that the inhomogeneity reduces as $N$ increases, as the fraction of atoms in the ``bulk'' vs the edges increases with $N$. This effect is more pronounced for the chain, where the fraction of bulk atoms scales as $1/N$, than the square array, where the fraction scales as $1/\sqrt{N}$, as can be seen in the inset to Supplementary Figure~\ref{SIFig4}(c).

\section*{Derivation of $\gtau$ for an imperfectly prepared initial state}

Here we consider the role of ``single-hole'' imperfections, i.e., where not all atoms are in the excited state. This state reads
\begin{equation}
\ket{\psi} = \sqrt{1 - \sum\limits_{a=1}^N|\zeta_a|^2}\bigotimes_{n=1}^N \ket{e}_n + \sum\limits_{a=1}^N \zeta_a \ket{g}_a \bigotimes_{n =1\neq a}^N \ket{e}_{n},\label{oneholestate}
\end{equation}
where $\zeta_a$ is the complex coefficient for the single-hole state in which atom $a$ is in the ground state.

The quantities required to calculate $\gtau$ do not mix states with different excitation numbers so we can evaluate the single-hole contribution separately to the fully-excited contribution. On the single-hole state, the expectation values required to calculate $\gtau$ are calculated as
\begin{subequations}
\begin{align}
&\left(\sum\limits_{a=1}^N \zeta_a^* \bra{g}_a \bigotimes_{n \neq a} \bra{e}_{n}\right) \heg^i\hge^j \left(\sum\limits_{b=1}^N \zeta_b \ket{g}_b \bigotimes_{p \neq b} \ket{e}_{p} \right) = \left(\delta_{ij}\delta_{ab} + \delta_{ib}\delta_{ja}\right)\left( 1 - \delta_{ia}\right)\\
&\left(\sum\limits_{a=1}^N \zeta_a^* \bra{g}_a \bigotimes_{n \neq a} \bra{e}_{n}\right) \heg^i\heg^j\hge^l\hge^m \left(\sum\limits_{b=1}^N \zeta_b \ket{g}_b \bigotimes_{p \neq b} \ket{e}_{p} \right) \notag\\
&\;\;\;\;\;\;\;\;\;\;\;\;\;\;\;\;= \left[\delta_{ab}\left(\delta_{il}\delta_{jm} + \delta_{im}\delta_{jl}\right) + \delta_{ib}\left(\delta_{jl}\delta_{ma} + \delta_{jm}\delta_{la}\right) + \delta_{jb}\left(\delta_{il}\delta_{ma} + \delta_{im}\delta_{la}\right)\right]\left(1-\delta_{ij}\right)\left(1-\delta_{ia}\right)\left(1-\delta_{ja}\right)
\end{align}
\end{subequations}

We calculate the numerator and denominator of $\gtau$ separately. The denominator is as follows
\begin{align}
\notag\sum\limits_{\nu=1}^N \Gamma_\nu \braket{\jop_\nu^\dagger \jop_\nu} &= \sum\limits_{\nu=1}^N \Gamma_\nu \sum\limits_{a,b,i,j=1}^N \zeta_a^*\zeta_b \alpha^*_{\nu,i}\alpha_{\nu,j} \left(\delta_{ij}\delta_{ab} + \delta_{ib}\delta_{ja}\right)\left(1 - \delta_{ia}\right) \\
\notag&= \sum\limits_{\nu=1}^N \Gamma_\nu \left[\sum\limits_{a,i=1}^N |\zeta_a|^2 |\alpha_{\nu,i}|^2 + \zeta_a^*\zeta_i \alpha^*_{\nu,i}\alpha_{\nu,a} - 2\sum\limits_{i=1}^N |\zeta_i|^2|\alpha_{\nu,i}|^2\right] \notag\\&= (N-2)\Gamma_0\sum\limits_{a=1}^N |\zeta_a|^2 + \sum\limits_{a,i,\nu=1}^N \Gamma_\nu\zeta_a^*\zeta_i \alpha^*_{\nu,i}\alpha_{\nu,a}.
\end{align}
Following a similar procedure, the numerator is readily found to be
\begin{align}
\notag&\sum\limits_{\nu,\mu=1}^N \Gamma_\nu\Gamma_\mu \braket{\jop_\nu^\dagger \jop_\mu^\dagger \jop_\mu \jop_\nu} = \sum\limits_{\nu,\mu=1}^N \Gamma_\nu\Gamma_\mu \sum\limits_{a,b,i,j,l,m=1}^N \zeta_a^*\zeta_b \alpha^*_{\mu,i}\alpha^*_{\nu,j}\alpha_{\nu,l}\alpha_{\mu,m}[\delta_{ab}\left(\delta_{il}\delta_{jm} + \delta_{im}\delta_{jl}\right) + \delta_{ib}\left(\delta_{jl}\delta_{ma} + \delta_{jm}\delta_{la}\right) \notag\\&\;\;\;\;\;\;\;\;\;\;\;\;\;\;\;\;\;\;\;\;\;\;\;\;\;\;\;\;\;\;\;\;\;\;\;\;\;\;\;\;\;\;\;+ \delta_{jb}\left(\delta_{il}\delta_{ma} + \delta_{im}\delta_{la}\right)]\left( 1- \delta_{ij} - \delta_{ia} - \delta_{ja} + 2\delta_{ij}\delta_{ia} \right) \notag\\
&= \sum\limits_{a=1}^N |\zeta_a|^2 \left[\left(N^2-6N+12\right)\Gamma_0^2  + \sum\limits_{\nu=1}^N \Gamma_\nu^2\left(1 - 4|\alpha_{\nu,a}|^2 \right)\right] + \sum\limits_{a,i,\nu=1}^N \zeta_a^*\zeta_i \left[ \left(2N-8\right)\Gamma_0\Gamma_\nu  + 2\Gamma_\nu^2\right]\alpha_{\nu,i}^* \alpha_{\nu,a}
\end{align}
We can now combine these with the fully-excited results to find $\gtau$ for the state given by Eq.~\eqref{oneholestate}
\begin{align}
&\gtau = \notag\\
&\frac{\left(N^2-2N\right)\Gamma_0^2 + \sum\limits_{\nu=1}^N \Gamma_\nu^2 - 4\sum\limits_{a=1}^N |\zeta_a|^2 \left[\left(N-3\right)\Gamma_0^2  + \sum\limits_{\nu=1}^N \Gamma_\nu^2|\alpha_{\nu,a}|^2 \right] + \sum\limits_{a,i=1}^N \zeta_a^*\zeta_i \left[ \sum\limits_{\nu=1}^N \left(\left(2N-8\right)\Gamma_0\Gamma_\nu  + 2\Gamma_\nu^2\right)\alpha_{\nu,i}^* \alpha_{\nu,a} \right]}{\left[\left(N - 2\sum\limits_{a=1}^N |\zeta_a|^2\right)\Gamma_0  + \sum\limits_{a,i,\nu=1}^N \Gamma_\nu\zeta_a^*\zeta_i\alpha^*_{\nu,i}\alpha_{\nu,a}\right]^2 }.\label{g2imperfect}
\end{align}

\begin{figure}[t!]
\includegraphics[width=0.7\textwidth]{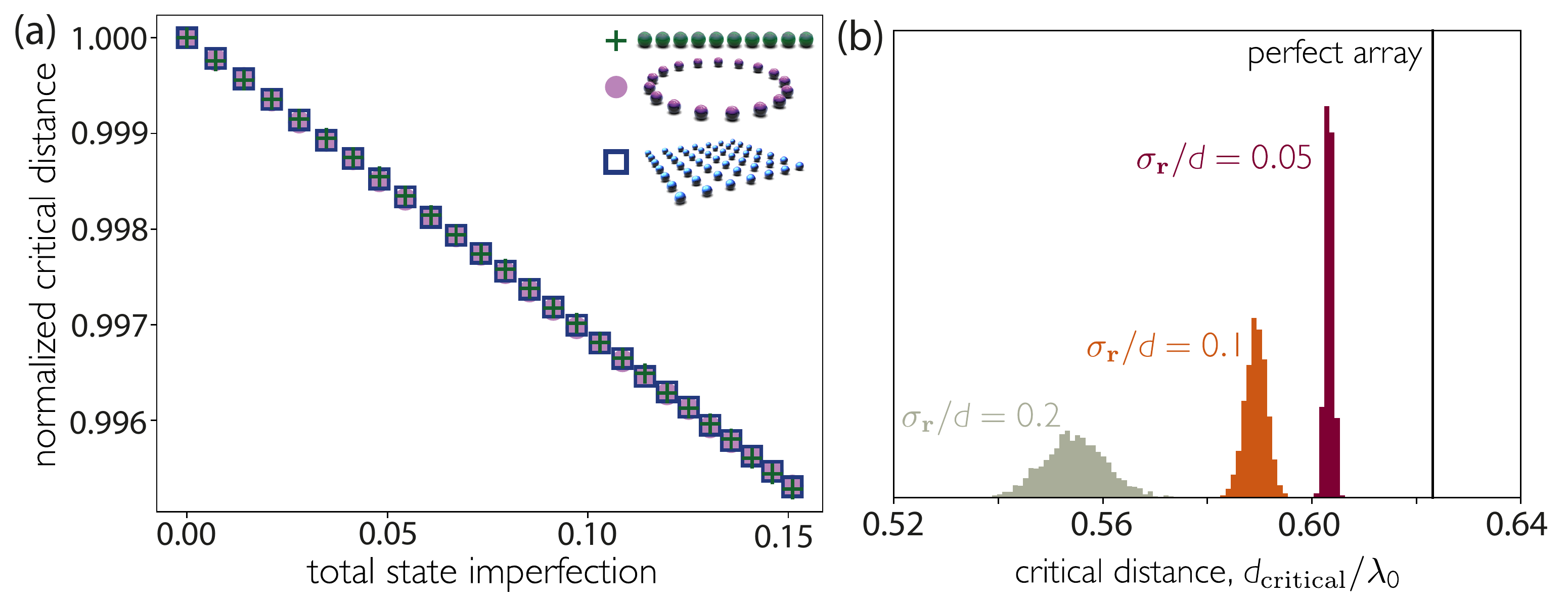}
\caption{(a) Impact of imperfections in the initial state on the critical distance at which superradiance disappears. The critical distance is found for different shape arrays of 36 atoms prepared in coherent spin states of the form given by Eq.~\eqref{CSS}, and plotted normalized by the critical distance for the fully inverted array. Atoms are arranged in the $x-y$ plane, or along the $x$-axis for the chain, with polarization axis along $z$ and drive along $x$. (b) Impact of classical spatial disorder on superradiance. The histogram shows the critical distance for 2000 configurations of a $12\times 12$ atom array with 3D Gaussian noise added to positions. Noise is added proportionally to the inter-atomic distance. Atoms are polarized perpendicular to the plane.}\label{SIFig1}
\end{figure}

To investigate the impact of the imperfect initial state, we consider coherent spin states of the form
\begin{equation}
    \ket{\varphi,\mathbf{k}} = \bigotimes\limits_{j=1}^N \left(\sqrt{1-\varphi} \ket{g}_j + \mathrm{e}^{i\mathbf{k}\cdot\mathbf{r}_j}\sqrt{\varphi} \ket{e}_j\right).
\end{equation}
These would be produced experimentally by a short, intense pulse of duration $\tau \ll\{ \left(N\Gamma_0\right)^{-1}, J_{12}^{-1}\}$. Here, we consider $\varphi \approx 1$ such that we truncate the state to the form (here left unnormalized for simplicity)
\begin{equation}
    \ket{\varphi,\mathbf{k}} \approx \sqrt{\varphi^N}\bigotimes\limits_{j=1}^N \ket{e}_j + \sum\limits_{j=1}^N  \mathrm{e}^{-i\mathbf{k}\cdot\mathbf{r}_j} \sqrt{\varphi^{N-1}(1-\varphi)} \ket{g}_j \bigotimes\limits_{l\neq j} \ket{e}_l \label{CSS}
\end{equation}
and use Eq.~\eqref{g2imperfect} to calculate the critical distance for imperfect initial states.  Supplementary Figure~\ref{SIFig1}(a) shows that the impact is marginal. For a total imperfection of $15\%$, the critical distance drops by a factor of only $0.4\%$. For these small imperfections in the initial state, the relative decrease in $\dcrit$ is approximately linear, and seems to be independent of the array geometry.

\section*{Critical distance in the presence of classical spatial disorder}

Supplementary Figure~\ref{SIFig1}(b) shows that superradiance is robust to classical disorder the position of in the emitters. We add a randomly-generated 3D Gaussian noise to each emitter position with standard deviation $\sigma_{\mathbf{r}}$ in all directions. We stochastically generate a large number of arrays and find the critical distance at which superradiance is lost. 

\section*{Considerations for solid-state emitters}

Solid-state emitters constitute an alternative platform for producing emitter arrays, as strongly sub-wavelength distances can be achieved simply through fabrication, without the need for optical trapping. However, these emitters have other issues that may negatively impact collective decay. Here, we consider the impact of inhomogeneous broadening and non-radiative decay.

\subsection*{Inhomogeneous broadening}

For non-identical emitters, we define each emitter to have frequency $\omega_0^i$ and spontaneous emission rate $\Gamma_0^i$, with mean values $\bar{\omega}_0$ and $\bar{\Gamma}_0$. If the frequency broadening is small, such that the spectral response is flat across the range of $\omega_0^i$, then frequency broadening does not impact the treatment of the dissipation and we can follow the derivation of $\gtau$ above with the alterations that the operator decay rates now obey
\begin{equation}
\sum\limits_{\nu=1}^N \Gamma_\nu |\alpha_{\nu,i}|^2 = \Gamma_0^i \;\;\;\; \mathrm{and} \;\;\;\; \sum\limits_{\nu=1}^N \Gamma_\nu = N\bar{\Gamma}_0.
\end{equation}
Therefore
\begin{align}
\gtau &= \frac{\sum\limits_{\nu,\mu=1}^N \Gamma_\nu\Gamma_\mu \left( 1 + \delta_{\nu\mu} - 2\sum\limits_{i=1}^N|\alpha_{\nu,i}|^2|\alpha_{\mu,i}|^2\right)}{N^2\bar{\Gamma}_0^2}= 1 + \sum\limits_{\nu=1}^N \left( \frac{\Gamma_\nu}{N\bar{\Gamma}_0}\right)^2 - 2\sum\limits_{i=1}^N \left(\frac{\Gamma_0^i}{N\bar{\Gamma}_0} \right)^2.
\end{align}
This can be recast in terms of two variances as
\begin{equation}
\gtau = 1 + \frac{1}{N}\left[\mathrm{Var}\left(\frac{\Gamma_\nu }{ \bar{\Gamma}_0}\right) - 1\right] -  \frac{2}{N}\mathrm{Var}\left(\frac{\Gamma_0^i }{ \bar{\Gamma}_0}\right).
\end{equation}
This expression is maximized for zero inhomogeneity, i.e. $\Gamma_0^i = \Gamma_0$, and so inhomogeneous broadening in the emitter decay rates strictly increases dephasing.

\subsection*{Non-radiative decay}

Solid-state emitters can decay without emitting light. We consider that this type of decay is not correlated (i.e., it is local). The master equation thus reads
\begin{equation}\label{masterequationnonrad}
\dot{\rho} = - \frac{\ii}{\hbar} [\mathcal{H},\rho] + \sum\limits_{\nu=1}^N \frac{\Gamma_\nu}{2} \left( 2\jop_\nu \rho \,\jop_\nu^\dagger - \rho\, \jop_\nu^\dagger\jop_\nu - \jop_\nu^\dagger\jop_\nu \rho \right) + \sum\limits_{i=1}^N \frac{\gamma_i}{2}\left(2 \hge^i \rho\, \heg^i - \rho\, \heg^i\hge^i - \heg^i\hge^i \rho \right),
\end{equation}
where $\gamma_i$ is the non-radiative decay rate of atom $i$.
We then write $\gtau$ as
\begin{equation}
\gtau = \frac{p(0,2)\sum\limits_{\nu,\mu=1}^N \Gamma_\nu\Gamma_\mu \braket{\jop_\nu^\dagger \jop_\mu^\dagger \jop_\mu \jop_\nu} +  \sum\limits_{i,\nu,\mu=1}^N \left[p_i(1,2)\Gamma_\nu\Gamma_\mu  \braket{\heg^i \jop_\nu^\dagger \jop_\mu^\dagger \jop_\mu \jop_\nu \hge^i} + p_i(2,2) \Gamma_\nu\Gamma_\mu  \braket{\jop_\nu^\dagger\heg^i \jop_\mu^\dagger \jop_\mu \hge^i \jop_\nu}\right]}
{\left(p(0,1)\sum\limits_{\nu=1}^N \Gamma_\nu \braket{\jop_\nu^\dagger \jop_\nu} + \sum\limits_{i,\nu=1}^N p_i(1,1)\Gamma_\nu \braket{\heg^i\jop_\nu^\dagger\jop_\nu\hge^i}\right)^2},
\end{equation}
where $p(0,j)$ is the probability of zero non-radiative events before the emission of $j$ photons, and $p_i(l,m)$ is the probability of a single non-radiative event occurring on atom $i$ right before the $m$th photon during the emission of $l$ photons. Terms with two or more non-radiative events are assumed to be negligible and are hence ignored, as we assume the non-radiative decay to be small, $\gamma_i \ll \Gamma_0$. 

We wish to expand $\gtau$ in the same manner as above, which requires the evaluation of the expectation values
\begin{subequations}
\begin{align}
\braket{\heg^i\heg^j\hge^l\hge^i} &= \delta_{jl}\left(1-\delta_{ij}\right), \\
\braket{\heg^i\heg^j\heg^l\hge^m\hge^n\hge^i} &= \left(\delta_{jm}\delta_{ln} + \delta_{jn}\delta_{lm}\right) \left(1-\delta_{jl}\right) \left(1-\delta_{ij}\right) \left(1-\delta_{il}\right), \label{eqp1}\\
\braket{\heg^j \heg^i \heg^l \hge^m \hge^i \hge^n} &= \left(\delta_{jm}\delta_{ln} + \delta_{jn}\delta_{lm}\right) \left(1-\delta_{jl}\right) \left(1-\delta_{ij}\right) \left(1-\delta_{il}\right).\label{eqp2}
\end{align}
\end{subequations}
By noting that Eqs.~\eqref{eqp1} and \eqref{eqp2} yield the same result, and substituting in the expressions for terms without non-radiative terms from above, we arrive to
\begin{equation}
\gtau = \frac{p(0,2)\left(N^2\bar{\Gamma}_0^2 + \sum\limits_{\nu=1}^N \Gamma_\nu^2 - 2\sum\limits_{i=1}^N \left(\Gamma^i_0\right)^2 \right) + \sum\limits_{i,\nu,\mu=1}^N \left[ p_i(1,2) + p_i(2,2)\right] \Gamma_\nu\Gamma_\mu  \braket{\heg^i \jop_\nu^\dagger \jop_\mu^\dagger \jop_\mu \jop_\nu \hge^i}}
{\left(p(0,1) N \Gamma_0 + \sum\limits_{i,\nu=1}^N p_i(1,1)\Gamma_\nu \braket{\heg^i\jop_\nu^\dagger\jop_\nu\hge^i}\right)^2}.
\end{equation}

We are interested in calculating $\gtau$ around the critical distance, where the second photon is emitted at approximately the same rate as the first, $N\Gamma_0$. In this situation, we can approximate the probabilities as
\begin{subequations}
\begin{align}
p(0,1) &= \frac{N\bar{\Gamma}_0}{N\bar{\Gamma}_0 + N\bar{\gamma}},\\
p_i(1,1) &= \frac{\gamma_i}{N\bar{\Gamma}_0 + N\bar{\gamma}},\\
p(0,2) &= \frac{N\bar{\Gamma}_0}{N\bar{\Gamma_0} + 2N\bar{\gamma}},\\
p_i(1,2) &= \frac{\gamma_i}{N\bar{\Gamma}_0 + 2N\bar{\gamma}},\\
p_i(2,2) &= \frac{\gamma_i}{N\bar{\Gamma}_0 + 2N\bar{\gamma}} = p_i(1,2),
\end{align}
\end{subequations}
where $\bar{\gamma}$ is the mean non-radiative decay rate. This approximation should also be valid for large $N$, where the emission of the first photon does not substantially alter the rate of the second photon. This simplifies the expression to
\begin{equation}
\gtau = \frac{p(0,2)\left(N^2\bar{\Gamma}_0^2 + \sum\limits_{\nu=1}^N \Gamma_\nu^2 - 2\sum\limits_{i=1}^N \left(\Gamma^i_0\right)^2 \right) + 2\sum\limits_{i,\nu,\mu=1}^N  p_i(1,2) \Gamma_\nu\Gamma_\mu  \braket{\heg^i \jop_\nu^\dagger \jop_\mu^\dagger \jop_\mu \jop_\nu \hge^i}}
{\left(p(0,1) N \Gamma_0 + \sum\limits_{i,\nu=1}^N p_i(1,1)\Gamma_\nu \braket{\heg^i\jop_\nu^\dagger\jop_\nu\hge^i}\right)^2}.
\end{equation}
We thus need to calculate
\begin{align}
&\sum\limits_{i,\nu=1}^N p_i(1,1)\Gamma_\nu \braket{\heg^i\jop_\nu^\dagger\jop_\nu\hge^i} = \sum\limits_{i,j,l,\nu=1}^N p_i(1,1) \Gamma_\nu \alpha^*_{\nu,j} \alpha_{\nu,l} \braket{\heg^i\heg^j\hge^l\hge^i}= \sum\limits_{i,j,l,\nu=1}^N p_i(1,1) \Gamma_\nu \alpha^*_{\nu,j} \alpha_{\nu,l} \delta_{jl}\left(1-\delta_{ij}\right) \notag \\
&= \sum\limits_{i,j,\nu=1}^N p_i(1,1) \Gamma_\nu |\alpha_{\nu,j}|^2 - \sum\limits_{i,\nu=1}^N p_i(1,1) \Gamma_\nu |\alpha_{\nu,i}|^2 = \left(N-1\right) \sum\limits_{i=1}^N p_i(1,1) \Gamma_0^i,
\end{align}
and
\begin{align}
&\sum\limits_{i,\nu,\mu=1}^N  p_i(1,2) \Gamma_\nu\Gamma_\mu  \braket{\heg^i \jop_\nu^\dagger \jop_\mu^\dagger \jop_\mu \jop_\nu \hge^i} = \sum\limits_{i,\nu,\mu=1}^N  p_i(1,2) \Gamma_\nu\Gamma_\mu \alpha_{\nu,j}^*\alpha_{\mu,l}^*\alpha_{\mu,m}\alpha_{\nu,n} \braket{\heg^i\heg^j\heg^l\hge^m\hge^n\hge^i} \notag\\
&= \sum\limits_{i,\nu,\mu=1}^N  p_i(1,2) \Gamma_\nu\Gamma_\mu \alpha_{\nu,j}^*\alpha_{\mu,l}^*\alpha_{\mu,m}\alpha_{\nu,n} \left(\delta_{jm}\delta_{ln} + \delta_{jn}\delta_{lm}\right) \left(1-\delta_{jl}\right) \left(1-\delta_{ij}\right) \left(1-\delta_{il}\right) \notag\\
&= \sum\limits_{i=1}^N p_i(1,2) \left[ N^2\bar{\Gamma}_0^2 +  \sum\limits_{\nu=1}^N \Gamma_\nu^2 \left(1 - 2|\alpha_{\nu,i}|^2\right) - 2N \Gamma_0^i \bar{\Gamma}_0 + 4 \left(\Gamma_0^i\right)^2 - 2\sum\limits_{j=1}^N \left(\Gamma_0^j\right)^2\right].
\end{align}

Combining these two expressions we obtain the second order correlation function near the critical distance as
\begin{align}
&\gtau = \\
&\frac{\left(\bar{\Gamma}_0 + 2\bar{\gamma}\right)N^2\bar{\Gamma}_0^2 + \sum\limits_{\nu=1}^N \Gamma_\nu^2 \left( \bar{\Gamma}_0 + 2\bar{\gamma} - \frac{4}{N} \sum\limits_{i=1}^N \gamma_i |\alpha_{\nu,i}|^2 \right) + \sum\limits_{i=1}^N \left(\frac{8\gamma_i}{N}-2\bar{\Gamma}_0\right) \left(\Gamma^i_0\right)^2  - \sum\limits_{i=1}^N 4\gamma_i \left( \Gamma_0^i \bar{\Gamma}_0 + \sum\limits_{j=1}^N \frac{\left(\Gamma_0^j\right)^2}{N}\right)}{\frac{\bar{\Gamma}_0 + 2\bar{\gamma}}{\left(N\bar{\Gamma}_0 + N\bar{\gamma}\right)^2} \left[ N^2\bar{\Gamma}_0^2 + (N-1)\sum\limits_{i=1}^N \gamma_i \Gamma_0^i \right]^2}. \notag
\end{align}
If each emitter has the same non-radiative decay rate $\gamma$, this simplifies to
\begin{align}
\gtau = \left(1 + \frac{\gamma}{\bar{\Gamma}_0}\right)^2\left(1  - \frac{4\gamma}{N\bar{\Gamma}_0+2N\gamma}\right)\frac{N^2\bar{\Gamma}_0^2 + \sum\limits_{\nu=1}^N \Gamma_\nu^2  -2\sum\limits_{i=1}^N  \left(\Gamma^i_0\right)^2}{\left[ N\bar{\Gamma}_0 + (N-1) \gamma \right]^2}. 
\end{align}

\subsection*{Superradiance with solid-state emitters}

\begin{figure}[h!]
    \centering
    \includegraphics[width=\textwidth]{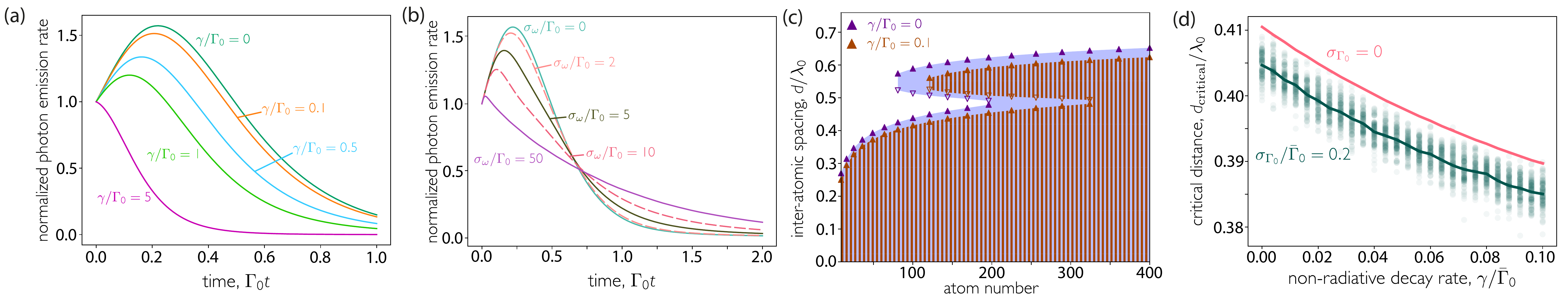}
    \caption{Impact of non-radiative decay and inhomogeneous broadening on superradiance. (a,b)~Photon emission rate from an initially inverted square array of $3\times3$ emitters and inter-atomic spacing $d=0.1\lambda_0$ in the presence of (a) non-radiative decay and (b) inhomogeneous broadening on the emitter resonance frequencies. In (b), plotted curves are the average of 100 stochastically generated instances with Gaussian distributed noise of width $\sigma_\omega$. (c)~Boundaries between the burst (colored) and no-burst (white) regions as a function of inter-particle distance $d$ and emitter number for square arrays with and without non-radiative decay. The symbols $\bigtriangleup$ and $\bigtriangledown$ represent points where, with decreasing $d$, $\gtau$ goes above and below unity, respectively. (d)~Critical distance for square arrays of $8\times8$ emitters as a function of non-radiative decay rate. In the presence of inhomogeneous broadening, the decay rate of each emitter is calculated as a random sample of a Gaussian distribution with mean $\Gamma_0$ and standard deviation $\sigma_{\Gamma_0}$. Circles represent individual stochastic samples, and the solid line shows the average of 100 samples. In all cases, emitters are polarized perpendicular to the array and are assumed to have the same non-radiative decay rate $\gamma$.}
    \label{SIFig3}
\end{figure}

Superradiance persists in the presence of non-radiative decay and inhomogeneous broadening. Supplementary Figure~\ref{SIFig3}(a) shows that the superradiant burst survives levels of non-radiative decay as large as those of radiative decay. Nevertheless, increased non-radiative decay rates enhance dephasing, eventually destroying superradiance as the emission pathways are dominated by non-radiative routes. As a result, the critical distance at which the superradiant burst disappears is shifted to smaller distances. Supplementary Figure~\ref{SIFig3}(b) shows that the superradiant burst survives inhomogeneous broadening on the emitter resonance frequency even at levels beyond 10 times the linewidth. The burst is diminished in size and duration, but not destroyed. Supplementary Figure~\ref{SIFig3}(c) shows that non-radiative decay always provides a stricter bound on superradiance, although the impact is relatively small if radiative decay is still the dominant decay mechanism. As the level of non-radiative decay increases, the critical distance decreases, as shown in Supplementary Figure~\ref{SIFig3}(d). The addition of inhomogeneous broadening on the atoms' linewidths results in a further small decrease in the critical distance.

\end{document}